\documentclass[aps,prd,twocolumn,groupedaddress,nofootinbib,longbibliography]{revtex4-1}

\usepackage{amssymb}
\usepackage{amsmath}
\usepackage[linktocpage=true]{hyperref}

\begin{document}

\title{Spinning particles in vacuum spacetimes of different curvature types:\\
       Natural reference tetrads, and massless particles}

\author{O. Semer\'ak}
\email[]{oldrich.semerak@mff.cuni.cz}

\affiliation{Institute of Theoretical Physics, Faculty of Mathematics and Physics, Charles University in Prague, Prague, Czech Republic}

\date{\today}

\begin{abstract}
In a previous paper, we considered the motion of massive spinning test particles in the ``pole-dipole" approximation, as described by the Mathisson--Papapetrou--Dixon (MPD) equations, and examined its properties in dependence on the spin supplementary condition. We decomposed the equations in the orthonormal tetrad based on the time-like vector fixing the spin condition and on the corresponding spin, while representing the curvature in terms of the Weyl scalars obtained in the Newman--Penrose (NP) null tetrad naturally associated with the orthonormal one; the projections thus obtained did not contain the Weyl scalars $\Psi_0$ and $\Psi_4$.

In the present paper, we choose the interpretation tetrad in a different way, attaching it to the tangent $u^\mu$ of the world-line representing the history of the spinning body. Actually {\em two} tetrads are suggested, both given ``intrinsically" by the problem and each of them incompatible with one specific spin condition. The decomposition of the MPD equation, again supplemented by writing its right-hand side in terms of the Weyl scalars, is slightly less efficient than in the massive case, because $u^\mu$ cannot be freely chosen (in contrast to $V^\mu$) and so the $u^\mu$-based tetrad is less flexible.

In the second part of this paper, a similar analysis is performed for massless spinning particles; in particular, a certain ``intrinsic" interpretation tetrad is again found. The respective decomposition of the MPD equation of motion is considerably simpler than in the massive case, containing only $\Psi_1$ and $\Psi_2$ scalars and not the cosmological constant. An option to span the spin-bivector eigen-plane, besides the world-line null tangent, by a main principal null direction of the Weyl tensor can lead to an even simpler result.
\end{abstract}


\maketitle

\section{Introduction}

In \cite{SemerakS-PRDI} (henceforth referred to as paper I), we studied the problem of motion of a massive spinning test particle (``pole-dipole" body) as described by the Mathisson--Papapetrou--Dixon (MPD) equations
\begin{align}
  \dot{p}^\mu &= -\frac{1}{2}\,{R^\mu}_{\nu\kappa\lambda}u^\nu S^{\kappa\lambda},
       \label{Papa-p} \\
  \dot{S}^{\alpha\beta} &= p^\alpha u^\beta-u^\alpha p^\beta,
       \label{Papa-S}
\end{align}
where
$p^\mu$ and $u^\mu$ denote the total momentum and four-velocity of the particle,
$S^{\mu\nu}$ is the particle-spin bivector, and
the dot denotes absolute derivative along $u^\mu$.
We restricted to {\em vacuum} space-times and focused on the dependence of the exercise on the spin supplementary condition $S^{\mu\nu}V_\nu=0$, necessary to fix ambiguity in the MPD equations, and on interpretation of the spin-curvature interaction in terms of the Weyl scalars. Starting from projection of the equations into a suitable orthonormal tetrad, we chose the latter's time vector to coincide with the ``reference observer" $V^\mu$ specifying the spin condition, and one of the spatial legs to be given by the spin vector $s^\mu$ connected with $S^{\mu\nu}$ by
$s^\mu:=-\frac{1}{2}\,\epsilon^{\mu\nu\rho\sigma}V_\nu S_{\rho\sigma}=-{^*\!}S^{\mu\nu}V_\nu$.
Rewriting the force term representing spin-curvature interaction in terms of the scalars $\Psi_{0\div 4}$, obtained by projecting the Weyl tensor onto the associated Newman--Penrose (NP) complex null tetrad, we found that the MPD-equation orthonormal-basis projections do not contain scalars $\Psi_0$ and $\Psi_4$. We then suggested a possible way how to choose the remaining two spatial basis vectors ``intrinsically", that is along directions provided by the geometry of the problem itself; this choice is applicable when $u^\mu\nparallel p^\mu$ (an alternative tetrad, usable in this situation -- but not together with the Mathisson-Pirani condition, on the contrary --, is added in the present paper, remark \ref{remark-to-paperI}).

In order to find how the problem looks in space-times of some particular curvature type, we aligned the first vector $k^\mu$ of the NP tetrad with the highest-multiplicity principal null direction (PND) of the Weyl tensor by a suitable choice of $V^\mu$, reproducing at the same time a given spin, either described by $s^\mu$ or $S^{\mu\nu}$ according to the MPD equations. More specifically, the plan goes like this: have a generic space-time (thus {\em some} $k^\mu$ and other PNDs) and a generic particle (with {\em some} spin vector $s^\mu$ or spin tensor $S^{\mu\nu}$ at a given point). Aligning the first real vector of the NP tetrad with $k^\mu$, its second real vector $l^\mu$ can always be chosen so as to satisfy the relation $s^\mu=\frac{s}{\sqrt{2}}\,(k^\mu-l^\mu)$,\footnote
{More precisely, it is only not possible if $k_\mu s^\mu=0$.}
or, respectively, as an eigen-vector of $S^{\mu\nu}$ independent of $k^\mu$; finally $V^\mu$ is retro-defined by $V^\mu=\frac{1}{\sqrt{2}}\,(k^\mu+l^\mu)$. Projecting the MPD equation of motion into the orthonormal tetrad involving {\em these} $V^\mu$ and $s^\mu/s$ as the 0th and 1st vectors (and completed by {\em some} orthonormal $e_{\hat{2}}^\mu$ and $e_{\hat{3}}^\mu$), one obtains equations (66)--(69) of paper I,
\begin{eqnarray}
  -V_\mu\,\dot{p}^\mu
     &=& -2s\,{\rm Im}\Psi_2\,u^{\hat{1}} \nonumber \\
     &{}& -s\,({\rm Im}\Psi_3-{\rm Im}\Psi_1)\,u^{\hat{2}} \nonumber \\
     &{}& -s\,({\rm Re}\Psi_3+{\rm Re}\Psi_1)\,u^{\hat{3}} \,,
     \label{V,dotp} \\
  e^{\hat{1}}_\mu\,\dot{p}^\mu
     &=& -2s\,{\rm Im}\Psi_2\,u^{\hat{0}} \nonumber \\
     &{}& -s\,({\rm Im}\Psi_3+{\rm Im}\Psi_1)\,u^{\hat{2}} \nonumber \\
     &{}& -s\,({\rm Re}\Psi_3-{\rm Re}\Psi_1)\,u^{\hat{3}} \,,
     \label{e1,dotp,paperI} \\
  e^{\hat{2}}_\mu\,\dot{p}^\mu
     &=& +s\left(2\,{\rm Re}\Psi_2-\frac{\Lambda}{3}\right)u^{\hat{3}} \nonumber \\
     &{}& -s\,({\rm Im}\Psi_3-{\rm Im}\Psi_1)\,u^{\hat{0}} \nonumber \\
     &{}& +s\,({\rm Im}\Psi_3+{\rm Im}\Psi_1)\,u^{\hat{1}} \,, \\
  e^{\hat{3}}_\mu\,\dot{p}^\mu
     &=& -s\left(2\,{\rm Re}\Psi_2-\frac{\Lambda}{3}\right)u^{\hat{2}} \nonumber \\
     &{}& -s\,({\rm Re}\Psi_3+{\rm Re}\Psi_1)\,u^{\hat{0}} \nonumber \\
     &{}& +s\,({\rm Re}\Psi_3-{\rm Re}\Psi_1)\,u^{\hat{1}} \,,
     \label{e3,dotp,paperI}
\end{eqnarray}
where $u^{\hat\alpha}$ denote the tetrad components of four-velocity.

It is known that -- with the exception of Petrov type III -- it is possible to rotate the null tetrad so as to become ``transverse" in the sense that the corresponding $\Psi_1$ and $\Psi_3$ projections vanish (instead of the usual elimination of $\Psi_0$ and $\Psi_4$). If such a rotation of the tetrad was feasible ({\em in addition} to the above), the spinning-particle motion would be fully determined by $\Psi_2$ and by the cosmological constant (because $\Psi_0$ and $\Psi_4$ are not involved from the beginning). Unfortunately, this could only be achieved by chance, because the necessary rotation involves all the NP vectors (in dependence on Weyl scalars in the original NP tetrad), in particular, it fixes the $(k^\mu,l^\mu)$ plane, so $l^\mu$ can{\em not} be chosen to lie in the $(k^\mu,s^\mu)$ plane at the same time.

In the last part of paper I, we discussed the implications of the spin conditions mostly considered in the literature, mainly advocating the condition $\dot{V}^\mu=0$ which leads to $u^\mu\parallel p^\mu$ and generalizing it, and finally checked several particular types of motion.

In the present paper, let us proceed in a similar way, but choosing a different orthonormal tetrad, namely the one tied to $u^\mu$ as the time vector. In section \ref{u-tetrad}, we suggest -- as a counter-part of the ``intrinsic" tetrad based on $V^\mu$ considered in paper I -- a $u^\mu$-based tetrad which follows naturally from geometry of the problem. If trying to adapt the interpretation tetrad to the Weyl-tensor PNDs, one is either lead to the situation when $S^{\mu\nu}u_\nu=0$, so the Mathisson-Pirani condition holds (thus returning to the respective section of paper I), or one has to release the ``natural" association of the orthonormal tetrad with the NP tetrad, namely to compute the Weyl scalars in a NP tetrad which is {\em not} naturally associated with the orthonormal tetrad into which the MPD equations have been projected. Both possibilities are worked out, with type-N space-time mentioned as an example. Implications of specific spin supplementary conditions are considered in section \ref{SSCs}, pointing out, in particular, that for $u^\mu\parallel p^\mu$ a different tetrad has to be devised since the original one degenerates (similarly as its counter-part employed in paper I).

In the second part (section \ref{massless}), we turn to spinning particles with zero rest mass. Starting from a summary of what is known from the literature, we study the geometry of the massless problem in a similar way as its ``massive" counter-part before. In particular, we again propose a certain natural NP tetrad and the associated orthonormal frame, which follow from the geometry of the problem itself, and inquire about the properties of the MPD equation of motion when projected there. Also the properties of the orthonormal frame are examined, including the circumstance $p^\mu\parallel k^\mu$ when the frame is not available (and $\dot{p}^\mu$ is itself proportional to the world-line null tangent $k^\mu$).

First, however, let us remind that the space-time is supposed to be {\em vacuum}, possibly involving a non-zero cosmological constant $\Lambda$, the metric signature is ($-$+++) and geometrized units are used in which $c=1$, $G=1$. Greek indices run 0--3, latin indices 1--3, and summation convention is followed. The dot denotes absolute derivative with respect to the particle's proper time $\tau$, the asterisk denotes Hodge dual and overbar indicates complex conjugation.
The Riemann tensor is defined by $V_{\nu;\kappa\lambda}-V_{\nu;\lambda\kappa}={R^\mu}_{\nu\kappa\lambda}V_\mu$ and the Levi-Civita tensor as
\begin{equation}
  \epsilon_{\mu\nu\rho\sigma}=\sqrt{-g}\,[\mu\nu\rho\sigma], \quad
  \epsilon^{\mu\nu\rho\sigma}=-\frac{1}{\sqrt{-g}}\,[\mu\nu\rho\sigma],
\end{equation}
where $g$ is the determinant of covariant metric and $[\mu\nu\rho\sigma]$ is the permutation symbol fixed by $[0123]:=1$.
Please, see (e.g.) paper I for an introductory summary on the spinning-particle problem, including basic as well as recent references.

\section{Vacuum MPD equations in a tetrad tied to $u^\mu$}
\label{u-tetrad}

The reference observer $V^\mu$, in terms of which the spin supplementary condition is written ($S^{\mu\sigma}V_\sigma=0$), can be chosen freely, so it is generically possible to attach it to a {\em given} NP tetrad by taking $V^\mu:=\frac{1}{\sqrt{2}}\,(k^\mu+l^\mu)$. This is {\em not} in general possible with $u^\mu$, because this has to be obtained from $p^\mu$ which in turn is determined by the MPD equations, so none of these two vectors can be {\em chosen}. Hence the procedure will have to be different, namely based on given $u^\mu$ and $k^\mu$.

We will again start from the MPD equation of motion, rewritten in terms of spin vector $s^\mu$ in the form (39) of paper I,
\begin{align}
  \dot{p}^\mu &= {^*\!R^\mu}_{\nu\alpha\beta}u^\nu s^\alpha V^\beta \nonumber \\
              &= \left({^*\!C^\mu}_{\nu\alpha\beta}+
                       \frac{\Lambda}{3}\,{\epsilon^\mu}_{\nu\alpha\beta}\right)
                 u^\nu s^\alpha V^\beta \,, \label{Dp,s}
\end{align}
where we have used the vacuum relation between the Riemann-tensor and Weyl-tensor left duals ${^*\!R^\mu}_{\nu\alpha\beta}$ and ${^*\!C^\mu}_{\nu\alpha\beta}$ (in a vacuum they equal the right duals). Two advantages of having $u^\mu$ as the time vector of the tetrad are obvious: first, similarly as $V^\mu$ (and $s^\mu$, which we used in paper I), the four-velocity $u^\mu$ appears on the r.h. side among the vectors on which the dual Riemann is projected; and second, the whole $\dot{p}^\mu$ is from the beginning orthogonal to $u^\mu$, so its ``zeroth" component in such a tetrad vanishes automatically. (Note that none of these properties hold for the third major ``time" vector of the exercise, $p^\mu$.) Now, however, the question arises: which spatial vectors should one add to $u^\mu$, in order to complete the basis. Generally, there are two possibilities: either to take some vectors provided ``intrinsically" by the $p^\mu$, $u^\mu$, $s^\mu$, $V^\mu$ geometry (possibly also including derivatives of these vectors), or to try to somehow connect the spatial basis directly to the curvature structure, while staying in a space orthogonal to $u^\mu$.

The first, ``intrinsic" possibility can be proposed in analogy with paper I. Actually, denoting
\[\gamma:=-u_\mu V^\mu \;(>0), \quad \mu:=-p_\mu V^\mu \;(>0),\]
we chose there the basis 
\begin{equation}
  V^\mu, \;\; s^\mu, \;\; \mu u^\mu-\gamma p^\mu, \;\; (s^2\delta^\mu_\nu-s^\mu s_\nu)\,\dot{V}^\nu
\end{equation}
(or rather its normalized version), made of the eigen-vectors $V^\mu$ and $s^\mu$ of the spin bivector $S_{\alpha\beta}=\epsilon_{\alpha\beta\mu\nu}V^\mu s^\nu$, and of the eigen-vectors $(\mu u^\mu-\gamma p^\mu)$ and $(s^2\delta^\mu_\nu-s^\mu s_\nu)\,\dot{V}^\nu$ of its dual ${^*\!}S^{\mu\nu}=s^\mu V^\nu-V^\mu s^\nu$.
As a counterpart of this basis, we suggested the quadruple made of $u^\mu$ and spatial vectors
\begin{align}
  & p^\mu-mu^\mu = -\dot{S}^{\mu\alpha}u_\alpha, \\
  & \gamma s^\mu+s_\nu u^\nu V^\mu = -{^*\!}S^{\mu\alpha}u_\alpha, \\
  & \epsilon^{\mu\iota\kappa\lambda}u_\iota(\gamma s_\kappa+s_\nu u^\nu V_\kappa)\,p_\lambda
    = -{^*\!}\dot{S}^{\mu\lambda}{^*\!}S_{\lambda\nu}u^\nu,
\end{align}
i.e. of the eigen-vectors $u^\mu$ and $(p^\mu-mu^\mu)$ (``hidden momentum") of the bivector ${^*\!}\dot{S}_{\mu\nu}=\epsilon_{\mu\nu\alpha\beta}p^\alpha u^\beta$, and of the eigen-vectors $(\gamma s^\mu+s_\nu u^\nu V^\mu)$ and $\epsilon^{\mu\iota\kappa\lambda}u_\iota(\gamma s_\kappa+s_\nu u^\nu V_\kappa)\,p_\lambda$ of the bivector $\dot{S}^{\alpha\beta}=p^\alpha u^\beta-u^\alpha p^\beta$.
In the above, $m$ is the particle mass with respect to $u^\mu$, given by $m:=-u_\mu p^\mu\,(>0)$.

Note that the last of the tetrad vectors can also be written in a different way: regarding the formula (see e.g. \cite{Hall-04}, equation (7.15))
\[{^*\!F}^{\mu\lambda}{^*\!H}_{\lambda\nu}=H^{\mu\lambda}F_{\lambda\nu}
  +\frac{1}{2}\,\delta^\mu_\nu\,F^{\alpha\beta}H_{\alpha\beta} \,,\]
valid for any two bivectors $F_{\mu\nu}$ and $H_{\mu\nu}$, we can rewrite
\begin{align}
  &\epsilon^{\mu\iota\kappa\lambda}u_\iota(\gamma s_\kappa+s_\nu u^\nu V_\kappa)\,p_\lambda= \nonumber\\
  &=-{^*\!}\dot{S}^{\mu\lambda}{^*\!}S_{\lambda\nu}u^\nu
   =-S^{\mu\lambda}\dot{S}_{\lambda\nu}u^\nu
    -\frac{1}{2}\,u^\mu\,\dot{S}^{\alpha\beta}S_{\alpha\beta}=  \nonumber \\
  &=-S^{\mu\lambda}\dot{S}_{\lambda\nu}u^\nu
    -u^\mu\epsilon^{\alpha\beta\gamma\delta}p_\alpha u_\beta V_\gamma s_\delta= \nonumber\\
  &=-S^{\mu\lambda}\dot{S}_{\lambda\nu}u^\nu-u^\mu s\dot{s}
   =-(\delta^\mu_\alpha+u^\mu u_\alpha)\,S^{\alpha\lambda}\dot{S}_{\lambda\nu}u^\nu=  \nonumber \\
  &=(\delta^\mu_\alpha+u^\mu u_\alpha)\,
    \epsilon^{\alpha\lambda\gamma\delta}V_\gamma s_\delta(p_\lambda-mu_\lambda) \,,
\end{align}
where we have used just basic forms of all the bivectors and relation (33) from paper I, i.e.
\begin{equation}  \label{sds=EsVup}
  s\dot{s} \equiv s_\mu\dot{s}^\mu
    = \frac{1}{2}\,S^{\alpha\beta}\dot{S}_{\alpha\beta}
    = S^{\alpha\beta}p_\alpha u_\beta
    = \epsilon^{\mu\nu\alpha\beta}s_\mu V_\nu u_\alpha p_\beta \,.
\end{equation}
Also, instead of the tetrad vectors $u^\mu$ and $(p^\mu-mu^\mu)$, it would be possible to use in the basis, for example, $p^\mu$ and $(mp^\mu-{\cal M}^2 u^\mu)$ (the latter being given by the component of $u^\mu$ orthogonal to $p^\mu$).

In order to make the tetrad orthonormal, one needs magnitudes of the spatial vectors:
\begin{align}
  &(p_\mu-mu_\mu)(p^\mu-mu^\mu)=m^2-{\cal M}^2, \\
  &(\gamma s_\mu+s_\nu u^\nu V_\mu)(\gamma s^\mu+s_\nu u^\nu V^\mu)=\gamma^2 s^2-(s_\nu u^\nu)^2, \\
  &\epsilon^{\mu\iota\kappa\lambda}u_\iota(\gamma s_\kappa+s_\nu u^\nu V_\kappa)\,p_\lambda\;
   \epsilon_{\mu\rho\sigma\tau}u^\rho(\gamma s^\sigma+s_\beta u^\beta V^\sigma)\,p^\tau= \nonumber\\
  &\quad ={^*\!}\dot{S}^{\mu\lambda}{^*\!}S_{\lambda\nu}u^\nu\,
           {^*\!}\dot{S}_{\mu\kappa}{^*\!}S^{\kappa\sigma}u_\sigma= \nonumber\\
  &\quad =-\frac{1}{2}\,\dot{S}^{\alpha\beta}\dot{S}_{\alpha\beta}
            {^*\!}S_{\lambda\nu}u^\nu{^*\!}S^{\lambda\sigma}u_\sigma= \nonumber\\
  &\quad =(m^2-{\cal M}^2)(\gamma s_\mu+s_\nu u^\nu V_\mu)(\gamma s^\mu+s_\nu u^\nu V^\mu)= \nonumber\\
  &\quad =(m^2-{\cal M}^2)\left[\gamma^2 s^2-(s_\nu u^\nu)^2\right].
\end{align}
Finally, regarding that the tetrad used in paper I was numbered as
\begin{align}
  e_{\hat{0}}^\mu &:= V^\mu,  \label{e0,paperI} \\
  e_{\hat{1}}^\mu &:= \frac{s^\mu}{s} \,, \\
  e_{\hat{2}}^\mu &:= \frac{\mu u^\mu-\gamma p^\mu}{\sqrt{\gamma^2(m^2-{\cal M}^2)-(\gamma m-\mu)^2}} \;, \\
  e_{\hat{3}}^\mu &:= \frac{\epsilon^{\mu\iota\kappa\lambda}V_\iota s_\kappa(\mu u_\lambda\!-\gamma p_\lambda)}
                           {s\,\sqrt{\gamma^2(m^2-{\cal M}^2)-(\gamma m-\mu)^2}} \;,
                      \label{e3,paperI}
\end{align}
let us do it similarly here,
\begin{align}
  e_{(0)}^\mu &:= u^\mu,  \label{e0} \\
  e_{(1)}^\mu &:= \frac{\gamma s^\mu+s_\nu u^\nu V^\mu}{\sqrt{\gamma^2 s^2-(s_\sigma u^\sigma)^2}} \;,
                  \label{e1} \\
  e_{(2)}^\mu &:= \frac{p^\mu-mu^\mu}{\sqrt{m^2-{\cal M}^2}} \;, \\
  e_{(3)}^\mu &:= \frac{\epsilon^{\mu\iota\kappa\lambda}
                        u_\iota(\gamma s_\kappa+s_\nu u^\nu V_\kappa)\,p_\lambda}
                       {\sqrt{m^2-{\cal M}^2}\,\sqrt{\gamma^2 s^2-(s_\sigma u^\sigma)^2}}
                  \label{e3}
\end{align}
(we distinguish the two tetrads by the different marking of their vector-numbering indices).

Clearly neither of the tetrads can be erected if $u^\mu\parallel p^\mu$ (see subsection \ref{our-condition} below).

\subsection{Basic observations}

One of the vectors we have proposed for the $u^\mu$-based tetrad, $(\gamma s^\mu+s_\nu u^\nu V^\mu)$, is a combination of $V^\mu$ and $s^\mu$, so it belongs to the eigen-plane of $S^{\mu\nu}$. If we select this plane to coincide with that spanned by the PND $k^\mu$ and a suitably chosen $l^\mu$, the vector $(\gamma s^\mu+s_\nu u^\nu V^\mu)$ will be linked with the curvature structure. This is actually the {\em best} what can be done in this respect; in particular, one cannot include in the basis {\em two} independent vectors lying in the $k^\mu$, $l^\mu$ plane, because it is impossible to make {\em both} of them orthogonal to $u^\mu$. Therefore, the above set of vectors seems to be a reasonable proposal from which to build a $u^\mu$-directed basis, which at the same time is attached to the curvature structure as closely as generically possible. (So far, however, the space-time is left completely general, and also the tetrad is not necessarily linked to the Weyl-tensor PNDs.)

Introducing the tetrad (\ref{e0})--(\ref{e3}), we can first write (\ref{Dp,s}) as
\begin{align}
  \dot{p}^\mu
  &= \frac{1}{\gamma}
     \left(\!{^*\!C^\mu}_{\nu\alpha\beta}+\frac{\Lambda}{3}\,{\epsilon^\mu}_{\nu\alpha\beta}\!\right)
     u^\nu (\gamma s^\alpha+s_\iota u^\iota V^\alpha) V^\beta  \label{Dp,s,tetrad} \\
  &= \frac{\sqrt{\gamma^2 s^2\!-\!(s_{(0)})^2}}{\gamma}
     \left(\!{^*\!C^\mu}_{(0)(1)(\delta)}+
           \frac{\Lambda}{3}\,{\epsilon^\mu}_{(0)(1)(\delta)}\!\right)
     V^{(\delta)},
\end{align}
where the relevant components of $V^{(\delta)}$ read
\begin{align}
  V^{(0)} := e^{(0)}_\mu V^\mu &= -u_\mu V^\mu \equiv \gamma \,,  \label{Ve0} \\
  V^{(2)} := e^{(2)}_\mu V^\mu &= \frac{\gamma m-\mu}{\sqrt{m^2-{\cal M}^2}} \;,  \label{Ve2} \\
  V^{(3)} := e^{(3)}_\mu V^\mu
          &= \frac{\gamma\,\epsilon^{\mu\iota\kappa\lambda}V_\mu u_\iota s_\kappa p_\lambda}
                  {\sqrt{m^2-{\cal M}^2}\,\sqrt{\gamma^2 s^2-(s_\sigma u^\sigma)^2}} \nonumber \\
          &= \frac{\gamma\,s_\mu\dot{s}^\mu}
                  {\sqrt{m^2-{\cal M}^2}\,\sqrt{\gamma^2 s^2-(s_\sigma u^\sigma)^2}}
             \label{Ve3}
\end{align}
(equation (\ref{sds=EsVup}) has been used).
It is clear that the cosmological constant does not occur in the $e^{(1)}_\mu\dot{p}^\mu$ component, i.e. in the  projection on $(\gamma s^\mu+s_\nu u^\nu V^\mu)$. Since the latter plays the role of spin in (\ref{Dp,s,tetrad}), this implies the same property we observed on $V^\mu$-tetrad decomposition in paper I: $\Lambda$ only influences motion in directions perpendicular to the spin.

When projecting $\dot{p}^\mu$ to the ``parenthesis" tetrad, one also notices that due to the orthogonality $u_\mu\dot{p}^\mu=0$ the ``second" component yields just
\[e^{(2)}_\mu\dot{p}^\mu=\frac{p_\mu\dot{p}^\mu}{\sqrt{m^2-{\cal M}^2}}
                        =\frac{-{\cal M}\dot{\cal M}}{\sqrt{m^2-{\cal M}^2}} \,,\]
where the mass ${\cal M}$ is given by ${\cal M}^2:=-p_\mu p^\mu\,(>0)$.

Let us also add some obvious identities useful when transforming between the ``hatted" and the ``parenthesized" tetrad":
\begin{align*}
  & \gamma\equiv -u_\mu V^\mu\equiv u^{\hat{0}}\equiv V^{(0)}, \\
  & \mu\equiv -p_\mu V^\mu\equiv p^{\hat{0}}, \\
  & m\equiv -u_\mu p^\mu\equiv p^{(0)}, \\
  & s_\mu u^\mu\equiv su^{\hat{1}}\equiv -s^{(0)}.
\end{align*}

\subsection{Decomposition in a curvature-adjusted tetrad. Which one?}

Employing Appendix A of paper I, where orthonormal components of the Weyl tensor (and consequently those of its dual) are expressed in terms of the $\Psi_0\div\Psi_4$ scalars, it is now easy to write down the decomposition of the MPD equation of motion (\ref{Dp,s}):
\begin{eqnarray}
  \frac{1}{\sigma}\,e^{(1)}_\mu\dot{p}^\mu
     &=&  -2\,{\rm Im}\Psi_2\,V^{(0)}      \nonumber \\
     &{}& -({\rm Im}\Psi_3+{\rm Im}\Psi_1)\,V^{(2)}     \nonumber \\
     &{}& -({\rm Re}\Psi_3-{\rm Re}\Psi_1)\,V^{(3)} \,,
     \label{e1,dotp} \\
  \frac{1}{\sigma}\,e^{(2)}_\mu\dot{p}^\mu
     &\equiv& \frac{-{\cal M}\dot{\cal M}}{\sigma\,\sqrt{m^2-{\cal M}^2}} \nonumber \\
     &=&  -({\rm Im}\Psi_3-{\rm Im}\Psi_1)\,V^{(0)}   \nonumber \\
     &{}& +\frac{1}{2}\,({\rm Im}\Psi_0-{\rm Im}\Psi_4)\,V^{(2)}  \nonumber \\
     &{}& +\left[{\rm Re}\Psi_2-\frac{1}{2}({\rm Re}\Psi_0+{\rm Re}\Psi_4)+\frac{\Lambda}{3}\right]
           V^{(3)},
     \label{e2,dotp} \\
  \frac{1}{\sigma}\,e^{(3)}_\mu\dot{p}^\mu
     &=&  -({\rm Re}\Psi_3+{\rm Re}\Psi_1)\,V^{(0)}  \nonumber \\
     &{}& -\left[{\rm Re}\Psi_2+\frac{1}{2}({\rm Re}\Psi_0+{\rm Re}\Psi_4)+\frac{\Lambda}{3}\right]
           V^{(2)}    \nonumber \\
     &{}& -\frac{1}{2}({\rm Im}\Psi_0-{\rm Im}\Psi_4)\,V^{(3)} \,,
     \label{e3,dotp}
\end{eqnarray}
where we abbreviated $\sigma:=\frac{\sqrt{\gamma^2 s^2-(s_\sigma u^\sigma)^2}}{\gamma}\,$.
Apparently the result is similar to the decomposition with respect to the $V^\mu$-based tetrad, given in equations (\ref{V,dotp})--(\ref{e3,dotp,paperI}), with one important difference: the components obtained in paper I do not contain $\Psi_0$ and $\Psi_4$, whereas now these scalars {\em are} present. On the other hand, the present approach has one big advantage: at any point, the reference observer $V^\mu$ can be chosen arbitrarily (in contrast to $u^\mu$), so one can in fact eliminate much of the above formulas.

Let us remind that the complex $\Psi$-scalars featuring in equations (\ref{e1,dotp})--(\ref{e3,dotp}) represent projections of the Weyl tensor onto the NP tetrad $(k^\mu,l^\mu,m^\mu,\bar{m}^\mu)$ naturally associated with its orthonormal counterpart (\ref{e0})--(\ref{e3}), namely connected with the latter by
\begin{align*}
  k^\mu:=\frac{1}{\sqrt{2}}\,(u^\mu+e^\mu_{(1)}), \quad &
  l^\mu:=\frac{1}{\sqrt{2}}\,(u^\mu-e^\mu_{(1)}), \\
  m^\mu:=\frac{1}{\sqrt{2}}\,(e^\mu_{(2)}+{\rm i}\,e^\mu_{(3)}), \quad &
  \bar{m}^\mu:=\frac{1}{\sqrt{2}}\,(e^\mu_{(2)}-{\rm i}\,e^\mu_{(3)}).
\end{align*}
One might also express the projections of the MPD equation onto the (\ref{e0})--(\ref{e3}) tetrad in terms of Weyl scalars obtained in some {\em different} NP tetrad, not associated with the given orthonormal tetrad, but then equations (\ref{e1,dotp})--(\ref{e3,dotp}) would look differently.

Consider now shortly our plan, i.e. tuning the tetrad to a given space-time curvature, similarly as in paper I. It will certainly be advantageous to identify the first vector $k^\mu$ of the NP tetrad with the Weyl-tensor PND of the highest multiplicity again. Should now the plane determined by $u^\mu$ and $k^\mu$ be made an eigen-plane of the spin bivector $S^{\mu\nu}$, one would have to resort to only one viable spin condition, with $V^\mu\equiv u^\mu$. This would however mean to return to paper I, section V.A, on MPD equations supplemented by the Mathisson--Pirani condition. Actually, setting $V^\mu=u^\mu$, one has $s_\sigma u^\sigma=0$, $\sigma=s$, $V^{(0)}=1$ and $V^{(i)}=0$, reducing the equations (\ref{e1,dotp})--(\ref{e3,dotp}) to
\begin{align}
  e^{(1)}_\mu\,\dot{p}^\mu &= -2s\,{\rm Im}\Psi_2 \,,                 \nonumber \\
  e^{(2)}_\mu\,\dot{p}^\mu &= -s\,({\rm Im}\Psi_3-{\rm Im}\Psi_1) \,, \nonumber \\
  e^{(3)}_\mu\,\dot{p}^\mu &= -s\,({\rm Re}\Psi_3+{\rm Re}\Psi_1) \,, \nonumber
\end{align}
which are just equations (97)--(99) of paper I.

If one insisted on the tight connection between the tetrad and the curvature structure, and at once on a sufficiently generic view (not pushing one into $V^\mu=u^\mu$), there is an alternative -- with $u^\mu$ used as the time vector of the orthonormal tetrad in which the MPD equations are decomposed, yet with the reference observer $V^\mu$ left free (for later adaptation of the NP tetrad to a given algebraic type). If adopting such a compromise, it is necessary to release the tight (``natural") connection between the NP tetrad and the orthonormal one. Specifically, one could consider instead the NP tetrad naturally associated with the {\em same} orthonormal tetrad as in paper I, i.e. with (\ref{e0,paperI})--(\ref{e3,paperI}).
Expressing such an alternative in another words, one could keep the NP tetrad (thus the Weyl scalars) from paper I, but decompose the MPD equations in the orthonormal tetrad (\ref{e0})--(\ref{e3}) instead of (\ref{e0,paperI})--(\ref{e3,paperI}). Rather than to derive such a ``hybrid" relations by transformation of the Weyl scalars, it is simpler to start from equations (\ref{V,dotp})--(\ref{e3,dotp,paperI}) and compose their new components according to transformation of the orthonormal basis. One finds easily that
\begin{align}
  e^{(1)}_\mu\dot{p}^\mu
    &= \frac{u^{\hat{0}}e^{\hat{1}}_\mu+u^{\hat{1}}V_\mu}
            {\sqrt{(u^{\hat{0}})^2-(u^{\hat{1}})^2}} \; \dot{p}^\mu \,, \\
  e^{(2)}_\mu\dot{p}^\mu
    &= -\,\sqrt{1-\frac{(\gamma m-\mu)^2}{\gamma^2(m^2-{\cal M}^2)}}
       \;\, e^{\hat{2}}_\mu\dot{p}^\mu \,.
\end{align}
To also find the ``hatted" decomposition of $e_{(3)}^\mu$, we recall $e_\mu^{(3)} V^\mu $ given in (\ref{Ve3}) and calculate the remaining components,
\begin{align*}
  e_\alpha^{(3)} e_{\hat{1}}^\alpha
    &= -\frac{u^{\hat{1}}\;e^{\hat{1}}_\alpha \dot{s}^\alpha}
             {\sqrt{m^2-{\cal M}^2}\;\sqrt{(u^{\hat{0}})^2-(u^{\hat{1}})^2}} \;, \\
  e_\alpha^{(3)} e_{\hat{2}}^\alpha
    &= 0 \,, \\
  e_\alpha^{(3)} e_{\hat{3}}^\alpha
    &= -\frac{1}
             {\sqrt{(u^{\hat{0}})^2-(u^{\hat{1}})^2}\;
              \sqrt{1-\frac{(\gamma m-\mu)^2}{\gamma^2(m^2-{\cal M}^2)}}} \;,
\end{align*}
which can then be inserted into
\[e^{(3)}_\mu\dot{p}^\mu
  = (e^{(3)}_\alpha e^\alpha_{\hat{0}})\,e^{\hat{0}}_\mu\dot{p}^\mu
    +(e^{(3)}_\alpha e^\alpha_{\hat{1}})\,e^{\hat{1}}_\mu\dot{p}^\mu
    +(e^{(3)}_\alpha e^\alpha_{\hat{3}})\,e^{\hat{3}}_\mu\dot{p}^\mu\,.\]
Since the decomposition (\ref{V,dotp})--(\ref{e3,dotp,paperI}) from paper I is expressed in terms of the ``hatted" four-velocity components, it is useful to add, as a counterpart of (\ref{Ve0})--(\ref{Ve3}), that
\begin{align}
  u^{\hat{0}} &\equiv V^{(0)} \equiv\gamma \,,  \\
  u^{\hat{1}} &= -s^{(0)}/s \,, \\
  u^{\hat{2}} &= \frac{\gamma m-\mu}
                      {\sqrt{\gamma^2(m^2-{\cal M}^2)-(\gamma m-\mu)^2}} \;, \\
  u^{\hat{3}} &= \frac{\gamma\,s_\mu\dot{s}^\mu}
                      {s\,\sqrt{\gamma^2(m^2-{\cal M}^2)-(\gamma m-\mu)^2}} \;.  \label{uhat3}
\end{align}
The last two components are proportional to $V^{(2)}$ and $V^{(3)}$, see (\ref{Ve2}) and (\ref{Ve3}), respectively.

\subsection{Algebraically special space-times: type-N example}

It is only meaningful to discuss the particular curvature types if one accepts the above compromise view, i.e. decomposes the MPD equations into the $u^\mu$-based orthonormal tetrad, but keeps the NP tetrad (in which $\Psi$-scalars are computed) unrelated, and thus free for adaptation to the curvature structure as in paper I. We saw above that one pays for this freedom by longer expressions for the MPD-equation projections. On the other hand, these equations ``inherit" from those obtained in paper I the lack of the $\Psi_0$ and $\Psi_4$ scalars.

For the most special Petrov type N, by using equations (82)--(83) of paper I, i.e.
\[-V_\mu\,\dot{p}^\mu\!=\!0, \;\;
  e^{\hat{1}}_\mu\,\dot{p}^\mu\!=\!0, \;\;
  e^{\hat{2}}_\mu\,\dot{p}^\mu
     \!=\!-\frac{\Lambda}{3}\,s\,u^{\hat{3}}, \;\;
  e^{\hat{3}}_\mu\,\dot{p}^\mu
     \!=\!\frac{\Lambda}{3}\,s\,u^{\hat{2}},\]
and (\ref{uhat3}) from above, we obtain
\begin{align}
  e^{(1)}_\mu\dot{p}^\mu &= 0 \,, \\
  e^{(2)}_\mu\dot{p}^\mu
    &= \frac{\Lambda}{3}\,s\,u^{\hat{3}}\,
       \sqrt{1-\frac{(\gamma m-\mu)^2}{\gamma^2(m^2-{\cal M}^2)}}=  \nonumber \\
    &= \frac{\Lambda}{3}\,\frac{s_\mu\dot{s}^\mu}{\sqrt{m^2-{\cal M}^2}} \;, \\
  e^{(3)}_\mu\dot{p}^\mu
    &= \frac{-\frac{\Lambda}{3}\,s\,u^{\hat{2}}}
            {\sqrt{(u^{\hat{0}})^2-(u^{\hat{1}})^2}\;
             \sqrt{1-\frac{(\gamma m-\mu)^2}{\gamma^2(m^2-{\cal M}^2)}}}=  \nonumber \\
    &= \frac{-\frac{\Lambda}{3}\,s^2\gamma\,(\gamma m-\mu)\sqrt{m^2-{\cal M}^2}}
            {\sqrt{\gamma^2 s^2-(s^{(0)})^2}\;
             [\gamma^2(m^2-{\cal M}^2)-(\gamma m-\mu)^2]} \;.
\end{align}
In the ``intrinsic" tetrad, we found, in equation (84) of paper I,
that ${\cal M}\dot{\cal M}=-(\Lambda/3)\,s_\mu\dot{s}^\mu$,
so we can also write the second equation as
\begin{equation}
    e^{(2)}_\mu\dot{p}^\mu
    = \frac{-{\cal M}\dot{\cal M}}{\sqrt{m^2-{\cal M}^2}} \;.
\end{equation}

The decomposition forms following for other Petrov types can also be obtained straightforwardly and we will not discuss them.

\section {Specific spin conditions}
\label{SSCs}

Let us briefly consider how the exercise looks when supplemented by the main spin conditions.
We will however not include the Mathisson--Pirani spin condition, $V^\mu\equiv u^\mu$, any more, because this simply reduces the problem to the form already treated in section V.A of paper I.

\subsection{Tulczyjew spin condition, $V^\mu\equiv p^\mu/{\cal M}$}
\label{T-condition}

We know from paper I (section V.B) that the Tulczyjew condition implies
$\gamma=m/{\cal M}$, $\mu={\cal M}$, $s_\mu p^\mu=0=s_\mu u^\mu$, $\dot{\cal M}=0$, $\dot{s}=0$ and $\sigma=s$,
so we have
\begin{equation}  \label{ealpha.V}
  V^{(0)}\equiv\gamma=\frac{m}{{\cal M}}\,, \;\;
  V^{(2)}=\frac{\sqrt{m^2-{\cal M}^2}}{{\cal M}}\,, \;\;
  V^{(3)}=0 \,,
\end{equation}
\begin{align}
  e^{(1)}_\mu\dot{p}^\mu &= \frac{s_\mu\dot{p}^\mu}{s}=e^{\hat{1}}_\mu\,\dot{p}^\mu \,, \\
  e^{(2)}_\mu\dot{p}^\mu &= 0 = e^{\hat{2}}_\mu\,\dot{p}^\mu \,, \\
  e^{(3)}_\mu\dot{p}^\mu &= \frac{\epsilon_{\mu\iota\kappa\lambda}
                                  \dot{p}^\mu p^\iota s^\kappa u^\lambda}
                                 {s\,\sqrt{m^2-{\cal M}^2}}
                          = \frac{{\cal M}\,S_{\mu\lambda}\dot{p}^\mu u^\lambda}
                                 {s\,\sqrt{m^2-{\cal M}^2}} = \nonumber \\
                         &= -\frac{{\cal M}\,\dot{S}_{\mu\lambda}p^\mu u^\lambda}
                                  {s\,\sqrt{m^2-{\cal M}^2}}
                          = \frac{\cal M}{s}\,\sqrt{m^2-{\cal M}^2}= \nonumber \\
                         &= e^{\hat{3}}_\mu\,\dot{p}^\mu \,,  \label{e3.dotp}
\end{align}
which reduces equations (\ref{e1,dotp})--(\ref{e3,dotp}) to
\begin{eqnarray}
  \frac{{\cal M}}{s^2}\,s_\mu\dot{p}^\mu
     &=&  -2m\;{\rm Im}\Psi_2     \nonumber \\
     &{}& -\sqrt{m^2-{\cal M}^2}\;({\rm Im}\Psi_3+{\rm Im}\Psi_1) \,,  \label{e1,dotp,Tulcz} \\
  0  &=&  -2m\,({\rm Im}\Psi_3-{\rm Im}\Psi_1)   \nonumber \\
     &{}& +\sqrt{m^2-{\cal M}^2}\;({\rm Im}\Psi_0-{\rm Im}\Psi_4) \,, \\
  \frac{{\cal M}^2}{s^2}
     &=&  -\frac{m\,({\rm Re}\Psi_3+{\rm Re}\Psi_1)}{\sqrt{m^2-{\cal M}^2}}  \nonumber \\
     &{}& -{\rm Re}\Psi_2-\frac{1}{2}({\rm Re}\Psi_0+{\rm Re}\Psi_4)-\frac{\Lambda}{3} \;.
     \label{e3,dotp,Tulcz}
\end{eqnarray}
So the projections of $\dot{p}^\mu$ into the $u^\mu$-based tetrad (``parenthesized") equal those into the $V^\mu$-based tetrad (``hatted"), and they appear somewhat simpler when written in terms of the $\Psi$-scalars computed in the null tetrad associated with the $V^\mu$-based orthonormal tetrad; this was done in paper I, equations (121)--(124):
\begin{eqnarray*}
  0 &=& s\,({\rm Im}\Psi_1-{\rm Im}\Psi_3) \,, \\
  e^{\hat{1}}_\mu\,\dot{p}^\mu
     &=& -\frac{2ms}{\cal M}\;{\rm Im}\Psi_2
         -s\,({\rm Im}\Psi_1+{\rm Im}\Psi_3)\,u^{\hat{2}} \,, \\
  0 &=& s\,({\rm Im}\Psi_1-{\rm Im}\Psi_3) \,, \\
  {\cal M}^2 &=& s^2\!\left(\!\frac{\Lambda}{3}\!-\!2\,{\rm Re}\Psi_2\!\right)
                 -\frac{ms^2\,({\rm Re}\Psi_1\!+\!{\rm Re}\Psi_3)}{\sqrt{m^2-{\cal M}^2}} \,.
\end{eqnarray*}
The reason for the difference is that when the above expression is written in terms of the Weyl scalars computed in the null tetrad associated with the $u^\mu$-based orthonormal tetrad, it contains, in addition to $\Psi_1$, $\Psi_2$ and $\Psi_3$, also $\Psi_0$ and $\Psi_4$.

\subsection{The condition $u^\mu\parallel p^\mu$: an alternative tetrad}
\label{our-condition}

If $u^\mu\parallel p^\mu$, then $p^\mu=mu^\mu$, $m={\cal M}$, $\dot{m}=\dot{\cal M}=0$, $\dot{p}^\mu=m\dot{u}^\mu$, $\dot{S}^{\mu\nu}=0$, ${^*\!}\dot{S}^{\mu\nu}=0$, $\dot{s}=0$, and $\mu=\gamma m$. The ``intrinsic" tetrad tied to $u^\mu$, (\ref{e0})--(\ref{e3}), cannot be used, because its last two vectors degenerate (the ``hidden momentum" vanishes).\footnote
{The tetrad (\ref{e0,paperI})--(\ref{e3,paperI}) suggested in paper I degenerates then in the same manner.}
One can however find a different orthonormal tetrad, usable even when $p^\mu=mu^\mu$ -- for example, one can choose, besides (\ref{e0}) and (\ref{e1}), a vector orthogonal to $u^\mu$, $V^\mu$ as well as $s^\mu$, i.e.
\[\epsilon^{\mu\nu\kappa\lambda}u_\nu V_\kappa s_\lambda=S^{\mu\nu}u_\nu \,,\]
and add the last one orthogonal to all $u^\mu$, $e_{(1)}^\mu$ and $S^{\mu\nu}u_\nu$
(we will number the two vectors in a reverse order):
\begin{align}
  e_{(2)}^\mu &:= \frac{\epsilon^{\mu\alpha\beta\gamma}u_\alpha(\gamma s_\beta+s_\rho u^\rho V_\beta)\,
                        \epsilon_{\gamma\nu\kappa\lambda}u^\nu V^\kappa s^\lambda}
                       {\sqrt{\gamma^2 s^2-(s_\sigma u^\sigma)^2}\,
                        \sqrt{(\gamma^2-1)s^2-(s_\sigma u^\sigma)^2}} = \qquad \nonumber \\
              & = \frac{(\delta^\mu_\nu+u^\mu u_\nu)(\gamma s^2 V^\nu+s_\rho u^\rho s^\nu)}
                       {\sqrt{\gamma^2 s^2-(s_\sigma u^\sigma)^2}\,
                        \sqrt{(\gamma^2-1)s^2-(s_\sigma u^\sigma)^2}} \;,  \label{e2*} \\
  e_{(3)}^\mu &:= \frac{S^{\mu\nu}u_\nu}{\sqrt{S^{\alpha\rho}u_\rho S_{\alpha\sigma}u^\sigma}}
                = \frac{\epsilon^{\mu\nu\kappa\lambda}u_\nu V_\kappa s_\lambda}
                       {\sqrt{(\gamma^2-1)s^2-(s_\sigma u^\sigma)^2}} \;.  \label{e3*}
\end{align}
Obviously, these vectors are not defined if the Mathisson--Pirani spin condition $S^{\mu\nu}u_\nu=0$ is applied.

The new basis vectors (\ref{e2*}) and (\ref{e3*}) provide -- independently of the spin condition -- projections
\begin{equation}
  V^{(2)}=\gamma\;\frac{\sqrt{(\gamma^2-1)s^2-(s_\rho u^\rho)^2}}
                       {\sqrt{\gamma^2 s^2-(s_\sigma u^\sigma)^2}} \;, \quad
  V^{(3)}=0 \,,
\end{equation}
so the equations (\ref{e1,dotp})--(\ref{e3,dotp}) assume a similar form as (\ref{e1,dotp,Tulcz})--(\ref{e3,dotp,Tulcz}), just with $e^{(2)}_\mu\dot{p}^\mu$ no longer equal to $\frac{-{\cal M}\dot{\cal M}}{\sqrt{m^2-{\cal M}^2}}$ and with slightly more complicated $\sigma$, $V^{(2)}$, $e^{(1)}_\mu\dot{p}^\mu$ and $e^{(3)}_\mu\dot{p}^\mu$.
Note that if employed together with the {\em Tulczyjew} condition, this alternative tetrad yields {\em exactly the same} projections of $V^\mu$ and $\dot{p}^\mu$ as the original tetrad, that is (\ref{ealpha.V})--(\ref{e3.dotp}).

In section V.C.1 of paper I we showed that the freedom which the condition $u^\mu\parallel p^\mu$ leaves to the choice of $V^\mu$ can be used to select the latter in such a manner that the corresponding spin $s^\mu$ is orthogonal to $u^\mu$ (thus also to $p^\mu$) and remains so along the whole trajectory. Specifically, this requires to select $V^\mu=u^\mu$ at some initial point and then to prescribe evolutions
\begin{equation}
  \dot{V}^\mu=\frac{\alpha}{\mu\,m^2}\,s^\mu,
  \qquad
  \dot{s}^\mu=\frac{\alpha\,s^2}{\mu\,m^2}\,V^\mu,
\end{equation}
with $\alpha$ given by
\[\alpha=\frac{{\cal M}^2}{s^2}\,\dot{p}^\mu s_\mu
        =\frac{{\cal M}^2}{s}\,e^{\hat{1}}_\mu\,\dot{p}^\mu \,.\]
Ensuring the above setting, one gets, at a generic point, the ``alternative" tetrad\footnote
{Let us stress that $s_\mu u^\mu=0$ does {\em not} in general mean $S_{\mu\nu}u^\mu=0$ (i.e. the Mathisson--Pirani condition): the spin bivector still has $V^\mu$ and $s^\mu$ as its eigen-vectors, while $u^\nu$ need not belong to the eigen-plane.}
\begin{align}
  & e_{(0)}^\mu = u^\mu, \qquad
    e_{(1)}^\mu = \frac{s^\mu}{s} \,, \nonumber \\
  & e_{(2)}^\mu = \frac{V^\mu-\gamma u^\mu}{\sqrt{\gamma^2-1}} \;, \quad
    e_{(3)}^\mu = \frac{S^{\mu\nu}u_\nu}{s\,\sqrt{\gamma^2-1}} \;,
\end{align}
and hence projections
\begin{equation}
  V^{(0)}=\gamma, \;\;
  V^{(1)}=0, \;\;
  V^{(2)}=\sqrt{\gamma^2-1} \,, \;\;
  V^{(3)}=0 \,.
\end{equation}
Consequently, equations (\ref{e1,dotp})--(\ref{e3,dotp}) reduce to
\begin{eqnarray}
  \frac{\alpha}{m^2}
     &=&  -2\gamma\,{\rm Im}\Psi_2
          -\sqrt{\gamma^2-1}\,({\rm Im}\Psi_3+{\rm Im}\Psi_1) \,,  \label{e1,dotp,OKS} \\
  \frac{V_\mu\dot{p}^\mu}{s(\gamma^2-1)}
     &=&  \frac{\gamma\,({\rm Im}\Psi_1-{\rm Im}\Psi_3)}{\sqrt{\gamma^2-1}}
          +\frac{1}{2}\,({\rm Im}\Psi_0-{\rm Im}\Psi_4) \,, \\
  \frac{S_{\mu\nu}u^\nu\dot{p}^\mu}{s^2(\gamma^2-1)}
     &=&  -\frac{\gamma\,({\rm Re}\Psi_3+{\rm Re}\Psi_1)}{\sqrt{\gamma^2-1}}  \nonumber \\
     &{}& -{\rm Re}\Psi_2-\frac{1}{2}({\rm Re}\Psi_0+{\rm Re}\Psi_4)-\frac{\Lambda}{3} \;.
     \label{e3,dotp,OKS}
\end{eqnarray}
This form is slightly more complicated than the Tulczyjew-condition counterpart (\ref{e1,dotp,Tulcz})--(\ref{e3,dotp,Tulcz}). Note that one cannot obtain the latter, or any other more special form, by resorting to $V^\mu\sim p^\mu$ or so, because by prescribing the initial value ($u^\mu$) and evolution of $V^\mu$, the reference observer was fixed and cannot be adjusted any more (it cannot be set proportional to $p^\mu$ or $u^\mu$, in particular).

\subsubsection{Remark: alternative to the ``intrinsic" tetrad of paper I}
\label{remark-to-paperI}

If $p^\mu=mu^\mu$, the tetrad (\ref{e0,paperI})--(\ref{e3,paperI}) employed in paper I is clearly meaningless as well. Let us suggest its substitute even usable in such a situation, thus supplementing paper I where we did not go into this detail. One of the vectors orthogonal to both $V^\mu$ and $s^\mu$ can obviously be chosen like above in the $u^\mu$-based tetrad case, namely according to (\ref{e3*}), and the last vector can be found in analogy with (\ref{e2*}), i.e. as the one orthogonal to all $V^\mu$, $s^\mu$ and (\ref{e3*}):
\begin{align}
  e_{(3)}^\mu &:= \frac{-S^{\mu\nu}u_\nu}{\sqrt{S^{\alpha\rho}u_\rho S_{\alpha\sigma}u^\sigma}}
                = \frac{\epsilon^{\mu\nu\kappa\lambda}V_\nu u_\kappa s_\lambda}
                       {\sqrt{(\gamma^2-1)s^2-(s_\sigma u^\sigma)^2}} \;, \\
  e_{(2)}^\mu &:= \frac{\epsilon^{\mu\alpha\beta\gamma}V_\alpha s_\beta\,
                        \epsilon_{\gamma\nu\kappa\lambda}V^\nu u^\kappa s^\lambda}
                       {s\,\sqrt{(\gamma^2-1)s^2-(s_\sigma u^\sigma)^2}} = \nonumber \\
              & = \frac{(\delta^\mu_\nu+V^\mu V_\nu-e^\mu_{\hat{1}}e_\nu^{\hat{1}})\,u^\nu}
                       {\sqrt{(\gamma^2-1)s^2-(s_\sigma u^\sigma)^2}} \;,
\end{align}
where we remind that $e^\mu_{\hat{1}}\equiv s^\mu/s$.
Therefore, the $e_{(2)}^\mu$ vector is represented by the component of $u^\mu$ orthogonal to both $V^\mu$ and $s^\mu$.
Again, this tetrad is not available if the Mathisson-Pirani condition holds, $S^{\mu\nu}u_\nu=0$.

\section{Massless particles}
\label{massless}

In paper I as well as so far here, we have been considering particles with non-zero rest mass. Let us now reserve some space to localized {\em massless} particles. It was shown by \cite{BailynR-77,BailynR-81} that the ``massless" situation, represented by a traceless energy-momentum tensor, implies\footnote
{In the massless case, let us use $k^\mu$ instead of $u^\mu$ for the tangent of the representative world-line $k^\mu$, while keeping the dot for covariant differentiation along that world-line, i.e. $\dot{X}:=X_{;\iota}k^\iota$.}
\begin{equation}
  m:=-p_\mu k^\mu=0 \quad (={\rm const~along} \; k^\mu);
\end{equation}
the same was also obtained by \cite{DuvalF-78} from conformal invariance of the action functional. Another results were that if $S_{\mu\nu}$ is space-like, $S_{\mu\nu}S^{\mu\nu}=:2s^2>0$,\footnote
{Otherwise it could hardly be understood as describing the rotational angular momentum.}
then
\begin{equation}
  S_{\mu\nu}k^\nu=0, \qquad
  k_\mu k^\mu=0, \quad
  \dot{k}^\mu\sim k^\mu,
\end{equation}
so -- as already suggested by \cite{Mashhoon-75} -- the Mathisson--Pirani condition automatically holds, and the particle follows a null geodesic. 

The MPD equations themselves (\ref{Papa-p}), (\ref{Papa-S}) remain the same,
\begin{equation}  \label{Papa,m=0}
  \dot{p}^\mu = -\frac{1}{2}\,{R^\mu}_{\nu\kappa\lambda}k^\nu S^{\kappa\lambda},  \qquad
  \dot{S}^{\alpha\beta} = p^\alpha k^\beta-k^\alpha p^\beta,
\end{equation}
yet one can only rarely take over results from the massive case (paper I) simply by putting $m=0$; namely, the assumption $u_\mu u^\mu=-1$ (and $V_\mu V^\mu=-1$) was used there frequently, whereas now $V^\mu\rightarrow u^\mu\rightarrow k^\mu$ turns out to be light-like.
From the second MPD equation, one sees immediately that the scalar $s$ called {\it helicity} is constant along $k^\mu$,
\[2s\dot{s}=S_{\mu\nu}\dot{S}^{\mu\nu}=0,\]
and that $\dot{S}^{\alpha\beta}$ is null since $\dot{S}^{\alpha\beta}\dot{S}_{\alpha\beta}=0$, with $k^\mu$ being a common eigen-vector of $S^{\alpha\beta}$ and $\dot{S}^{\alpha\beta}$.

Let us stop at $k^\mu$ for a while: here it represents the world-line tangent, while in paper I we denoted by $k^\mu$ the first vector of the Newman--Penrose (NP) interpretation tetrad. However, the tetrad was chosen so that $k^\mu$ (as well as its second vector $l^\mu$) lied in the eigen-plane of $S_{\mu\nu}$, which is just consistent with the present notation since the spin condition $S^{\mu\nu}k_\nu=0$ now necessarily holds, so $k^\mu$ is naturally taken as the main vector of the interpretation tetrad.

Multiplication of the second of MPD equations (\ref{Papa,m=0}) by $p_\beta$ and by $\dot{p}_\beta$ yields
\begin{align}
  {\cal M}^2 k^\alpha &= -\dot{S}^{\alpha\beta}p_\beta \,, \\
  {\cal M}\dot{\cal M}\,k^\alpha = (p_\mu\dot{p}^\mu)\,k^\alpha
                      &= -\dot{S}^{\alpha\beta}\dot{p}_\beta \,, \label{MdotM}
\end{align}
from where one sees that
\[\dot{S}^{\alpha\beta}\dot{p}_\beta = \ddot{S}^{\alpha\beta}p_\beta \,,
  \quad
  \dot{\cal M}\,\dot{S}^{\alpha\beta}p_\beta = {\cal M}\,\dot{S}^{\alpha\beta}\dot{p}_\beta \,.\]
Above, we have introduced $p_\mu p^\mu=:{\cal M}^2$ as in the massive case, but with a different ({\em plus}) sign -- we will see below that $p^\mu$ is space-like now!

Another difference from the massive case is that the spin vector defined analogously as there, by projection of the spin-bivector dual onto $V^\mu\rightarrow k^\mu$, is also null ($s_\mu s^\mu=0$) and proportional to $k^\mu$, 
\begin{equation}  \label{spin-vector}
  s^\mu := -\frac{1}{2}\,\epsilon^{\mu\nu\rho\sigma}k_\nu S_{\rho\sigma}
         = -{^*\!}S^{\mu\nu}k_\nu
         = s\,k^\mu \,.
\end{equation}
The null character of $s^\mu$ is seen immediately: $s_\mu s^\mu$ only contains terms involving $S_{\rho\sigma}k^\sigma=0$ or $k_\nu k^\nu=0$. The second claim, $s^\mu=sk^\mu$, follows from the fact that two real null vectors are orthogonal if and only if they are proportional to each other.
The above result also implies that $s^\mu$ parallel transports along $k^\mu$, specifically, if $k^\mu$ is affinely parametrized ($\dot{k}^\mu=0$), then
\begin{equation}
  \dot{s}^\mu=\dot{s}\,k^\mu+s\,\dot{k}^\mu=0.
\end{equation}

Once knowing that the particle moves on a geodesic and that its spin is proportional to the latter's tangent $k^\mu$, one might have little reason to continue the study, because the momentum $p^\mu$ is a ``strange thing" (space-like) anyway, so there is actually no demand to interpret its evolution $\dot{p}^\mu$. However, we show below that even in the massless case there naturally follows a (time-like) ``reference observer" and an associated (space-like) spin vector (whether the former is taken as primary or the latter), i.e. quantities which have the same meaning as in the massive-particle case and which are worth further consideration. We first realize that the null version of the Mathisson-Pirani condition leaves more freedom to the spin bivector than the time-like version, and then fix the remaining freedom by determining the remaining independent dimension of the spin-bivector eigen-plane. In doing so, we naturally introduce the reference observer $V^\mu$ and the corresponding spin $S^\mu$, and also note that one can in fact take advantage of this freedom and adjust the spin eigen-plane so as to contain a desired direction (independent of $k^\mu$), in particular the main PND of the host space-time.

\subsection{Null spin condition: $S^{\mu\nu}k_\nu=0$, $k_\mu k^\mu=0$}
\label{null-condition}

As reminded by \cite{BiniCGJ-06} in their treatment of massless spinning particles, the null version of the spin condition is less restricting than the ``full" time-like case. Generally speaking, vanishing of the projection of an object onto a {\em null} direction $k^\mu$ does {\em not} exclude that the object has a component proportional to $k^\mu$. In the case of our bivector $S^{\mu\nu}$, the ``time-like" condition $S^{\mu\nu}V_\nu=0$, considered in paper I, strictly determined its eigen-plane and blade, in particular, it implied that the bivector must read
\[S_{\mu\nu}=\epsilon_{\mu\nu\kappa\lambda}V^\kappa s^\lambda
            =-s\,\epsilon_{\mu\nu\kappa\lambda}k^\kappa l^\lambda
            ={\rm i}\,s\,m_\mu\wedge\bar{m}_\nu \,,\]
where the real null vectors $k^\mu$ and $l^\mu$ were related to $V^\mu$ and $s^\mu$ by
\[k^\mu=\frac{1}{\sqrt{2}}\left(V^\mu+\frac{s^\mu}{s}\right), \qquad
  l^\mu=\frac{1}{\sqrt{2}}\left(V^\mu-\frac{s^\mu}{s}\right),\]
and $m^\mu$ and $\bar{m}^\mu$ are complex null vectors (mutual complex conjugates) orthogonal to both $k^\mu$ and $l^\mu$ and normalized to $m_\mu\bar{m}^\mu=1$.
In contrast, the condition $S^{\mu\nu}k_\nu=0$ admits a more general form
\begin{equation}
  S_{\mu\nu}={\rm i}\,s\,m_\mu\wedge\bar{m}_\nu+k_\mu\wedge(L m_\nu+\bar{L}\bar{m}_\nu),
\end{equation}
where $m^\mu$ and $\bar{m}^\mu$ are some complex null vectors orthogonal to $k^\mu$ and normalized to $m_\mu\bar{m}^\mu=1$, and $L$ denotes an (arbitrary) magnitude of ``the other" independent spin component.
Speaking more generally, the spin vector (\ref{spin-vector}) follows uniquely from a {\em known} bivector, but the converse is not true: the bivector is not fully determined by the spin vector.

However, a simple non-null bivector has the whole {\em plane} of eigen-directions (with zero eigen-value), so there exists (or can be chosen) a second null direction $l^\mu$, independent of $k^\mu$, which is also ``annihilated", $S^{\mu\nu}l_\nu=0$. Provided it is normalized so that $k_\mu l^\mu=-1$, the conditions $S^{\mu\nu}k_\nu=0$ and $S^{\mu\nu}l_\nu=0$ require $L=-{\rm i}\,s\,\bar{m}_\nu l^\nu$ (ergo $\bar{L}={\rm i}\,s\,m_\nu l^\nu$). For the eigen-directions $k^\mu$ and $l^\mu$ known/chosen, the bivector is already determined uniquely (and it is possible to choose $m^\mu$ and $\bar{m}^\mu$ perpendicular to both, making $L=0$). Clearly, if there is some privileged null direction in space-time (call it $l^\mu$), one can take advantage of the freedom still remaining in the spin bivector subjected to the null condition $S^{\mu\nu}k_\nu=0$, and require that it also satisfy $S^{\mu\nu}l_\nu=0$, thus inclining the bivector's eigen-plane in the desired way.

\subsection{Spin-bivector eigen-plane}

Having introduced $l^\mu$ as the second independent eigen-vector of the spin bivector, we can multiply by $l_\beta$ the second equation of (\ref{Papa,m=0}), to get
\begin{equation}  \label{p-k,relation}
  p^\alpha = -p^\beta l_\beta\,k^\alpha + p_\perp^\alpha
\end{equation}
as a counter-part of equation
$\gamma\,p^\alpha=\mu\,u^\alpha+S^{\alpha\beta}\dot{V}_\beta$
which was numbered (21) in paper I.
We have introduced
\begin{align}
  S^{\alpha\beta}\dot{l}_\beta
    &= -\dot{S}^{\alpha\beta}l_\beta = p^\alpha+k^\alpha l_\beta p^\beta =  \nonumber \\
    &= (\delta^\alpha_\beta+k^\alpha l_\beta+l^\alpha k_\beta)\,p^\beta =: p_\perp^\alpha
\end{align}
as the part of $p^\mu$ orthogonal to the plane $(k^\mu,l^\mu)$;
it is a counter-part of the ``hidden momentum"
\[p^\alpha_{\rm hidden}:=(\delta^\alpha_\beta+u^\alpha u_\beta)\,p^\beta
                        =p^\alpha-mu^\alpha
                        =-\dot{S}^{\alpha\beta}u_\beta\]
from the massive case.

As already suggested above, we will use in the next section the NP tetrad based on independent real null vectors $k^\mu$ and $l^\mu$ which are both annihilated by $S_{\mu\nu}$ and which are normalized as $k_\mu l^\mu=-1$. Being null, $l^\mu$ certainly satisfies $\dot{l}^\mu l_\mu=0$, and, if the particle's geodesic is affinely parametrized ($\dot{k}^\mu=0$), $\dot{l}^\mu k_\mu=0$ as well, but $l^\mu$ need not be parallel along $k^\mu$ (i.e., $\dot{l}^\mu\neq 0$ in general). Actually, with helicity $s$ known, one can ``reconstruct" the spin bivector (and its dual) by
\begin{equation}  \label{bivectors,skl}
  S_{\alpha\beta}=-s\,\epsilon_{\alpha\beta\gamma\delta}k^\gamma l^\delta, \quad
  {^*\!}S^{\mu\nu}=s\,(k^\mu l^\nu-l^\mu k^\nu) \,.
\end{equation}
Multiplying the derivative
\begin{equation}  \label{dotS-k,dotl}
  \dot{S}_{\alpha\beta}=-s\,\epsilon_{\alpha\beta\gamma\delta}k^\gamma\dot{l}^\delta
\end{equation}
by $\epsilon^{\mu\nu\alpha\beta}l_\nu$, we have
\begin{equation}  \label{dotl.dotl}
  s^2\dot{l}^\mu=-S^{\mu\alpha}p_\alpha
  \quad\! \Rightarrow \quad\!
  s^2\dot{l}^\mu\dot{l}_\mu=\dot{S}^{\mu\alpha}l_\mu p_\alpha={\cal M}^2.
\end{equation}

We have again used $p^\alpha p_\alpha=:{\cal M}^2$, so with the sign different from the massive case.
Namely, $\dot{l}^\mu$ is clearly orthogonal to both $k^\mu$ and $l^\mu$ which span the eigen-plane of $S^{\mu\nu}$; and this eigen-plane is time-like by assumption, so $\dot{l}^\mu$ has to be space-like, hence ${\cal M}^2>0$. Besides, $\dot{l}^\mu$ is also seen to be orthogonal to $p^\mu$; the reason cannot (in general) be that $p^\mu$ also belongs to the eigen-plane of $S^{\mu\nu}$, because this would mean $\dot{l}^\mu=0$, so $\dot{S}^{\mu\nu}=0$ and, consequently, $p^\mu\parallel k^\mu$, which is not in general consistent with the MPD equation for $\dot{p}^\mu$ (cf. \cite{BailynR-81}, section V, and section \ref{p||k} below). Therefore, in generic situation the vectors $k^\mu$, $l^\mu$ and $p^\mu$ are independent.

Note that one learns from (\ref{bivectors,skl}) that $k^\mu$ is also annihilated by
\begin{equation}
  {^*\!}\dot{S}^{\mu\nu}=s\,(k^\mu\dot{l}^\nu-\dot{l}^\mu k^\nu),
\end{equation}
so it is {\em the} common null eigen-vector of $\dot{S}^{\mu\nu}$ and ${^*\!}\dot{S}^{\mu\nu}$. This confirms that $\dot{S}^{\mu\nu}$ is null and thus ${^*\!}\dot{S}^{\mu\nu}\dot{S}_{\alpha\nu}=0$ like in the massive case, similarly as ${^*\!}S^{\mu\nu}S_{\alpha\nu}=0$.

\subsection{Summary of eigen-vectors of the spin bivectors}
\label{eigen-vectors}

It is very easy now to summarize the independent eigen-vectors of all the bivectors involved.
The eigen-plane of $S^{\mu\nu}$ is time-like and it is spanned by $k^\mu$ and $l^\mu$, while
the eigen-plane of $\dot{S}^{\mu\nu}$ is null and spanned by $k^\mu$ and $\dot{l}^\mu$
(the two eigen-planes intersect ``along" $k^\mu$).
The eigen-vectors of ${^*\!}\dot{S}^{\mu\nu}$ are $k^\mu$ and $p^\mu$, their plane being null (because ${^*\!}\dot{S}^{\mu\nu}$ is null, as ``inherited''from $\dot{S}^{\mu\nu}$).
The last bivector, ${^*\!}S^{\mu\nu}$, is the only one which does not annihilate $k^\mu$, but clearly this is true for $\dot{l}^\mu$, while its second eigen-vector can be found among projections of $S^{\mu\nu}$; in particular, $S^{\mu\nu}\dot{l}_\nu\equiv p_\perp^\mu$ is certainly independent of $\dot{l}^\mu$. Both $\dot{l}^\mu$ and $p_\perp^\mu$ are space-like, as well as the eigen-plane spanned by them (this is confirmed by the time-like character of the dual spin bivector, ${^*\!}S^{\mu\nu}{^*\!}S_{\mu\nu}\!=\!-2s^2$).

Therefore, the massless case differs from the massive one in the null character of $\dot{S}^{\mu\nu}$ and ${^*\!}\dot{S}^{\mu\nu}$ (for a massive particle, $\dot{S}^{\mu\nu}$ is time-like and ${^*\!}\dot{S}^{\mu\nu}$ is (thus) space-like).

\subsection{A natural tetrad}

In section III.D of paper I, we suggested a natural orthonormal tetrad which is provided ``intrinsically", by geometry of the spinning-particle problem itself. In case of the Mathisson--Pirani supplementary condition, it was given by $u^\mu$, $s^\mu$ (the eigen-vectors of the spin bivector), by hidden momentum $(p^\mu-mu^\mu)$ and the vector product of the three. The vectors
\begin{equation}
  k^\mu, \;\; l^\mu; \quad p_\perp^\mu, \;\; \dot{l}^\mu
\end{equation}
we listed in the previous subsection can be used as such a natural tetrad here in the massless case. Actually, $k^\mu$ and $l^\mu$ span the (time-like) eigen-plane of $S^{\mu\nu}$, and $p_\perp^\mu$ with $\dot{l}^\mu$ span the space-like plane orthogonal to it, being orthogonal to each other as well. The first two, null vectors are normalized by $k_\mu l^\mu=-1$, and the second, space-like couple is seen immediately to have norms given by
\[p_\perp^\mu p^\perp_\mu = p^\mu p_\mu = {\cal M}^2 \,, \qquad
  \dot{l}^\mu\dot{l}_\mu = \frac{{\cal M}^2}{s^2} \,.\]
Needless to say, the space-like basis vectors
\begin{equation}  \label{e2,e3}
  e^\mu_{(2)}:=\frac{p_\perp^\mu}{\cal M} \,, \qquad
  e^\mu_{(3)}:=\frac{s\,\dot{l}^\mu}{\cal M}
\end{equation}
can be transformed into null ones by
\begin{equation}
  m^\mu:=\frac{1}{\sqrt{2}}\,(e^\mu_{(2)}+{\rm i}\,e^\mu_{(3)}), \quad
  \bar{m}^\mu:=\frac{1}{\sqrt{2}}\,(e^\mu_{(2)}-{\rm i}\,e^\mu_{(3)}),
\end{equation}
to complete the NP null tetrad to $(k^\mu,l^\mu,m^\mu,\bar{m}^\mu)$.

\subsection{Vacuum MPD equations in a natural tetrad}

Regarding that the spin condition $S^{\mu\nu}k_\nu=0$ holds, we naturally tie the interpretation tetrad to $k^\mu$. Proceeding as above, one assumes that $S^{\mu\nu}$ is space-like ($S_{\mu\nu}S^{\mu\nu}=2s^2>0$), which implies that it has a time-like eigen-plane. Within such a plane, it is possible to find {\em two} independent null eigen-vectors. Denote by $l^\mu$ ``the other one", independent of $k^\mu$, and normalize it by $k_\mu l^\mu=-1$. To complete the standard NP null tetrad, add two complex null vectors $m^\mu$ and $\bar{m}^\mu$, orthogonal to both $k^\mu$ and $l^\mu$ and normalized as $m_\mu\bar{m}^\mu=1$.

The MPD equation of motion (\ref{Papa,m=0}) for the massless case can now be written as
\begin{align}
  \dot{p}^\mu &=-\frac{1}{2}\,{R^\mu}_{\nu\kappa\lambda}k^\nu S^{\kappa\lambda}
               = \frac{s}{2}\,g^{\mu\rho}R_{\rho\nu\kappa\lambda}
                 \epsilon^{\kappa\lambda\alpha\beta}k^\nu k_\alpha l_\beta
                 \nonumber \\
              &= s\,g^{\mu\rho}\,{^*\!}R_{\rho\nu\alpha\beta}k^\nu k^\alpha l^\beta
               = -s\,{^*\!C^\mu}_{\nu\alpha\beta}k^\nu l^\alpha k^\beta,
  \label{*Cklk}
\end{align}
where $R^*_{\rho\nu\alpha\beta}$ and ${^*\!R}_{\rho\nu\alpha\beta}$ are the Riemann-tensor right and left duals (as in paper I, equation (39), we have used that they are equal in the vacuum case; this does not depend on the value of cosmological constant). Since
${^*\!R^\mu}_{\nu\alpha\beta}={^*\!C^\mu}_{\nu\alpha\beta}+\frac{\Lambda}{3}\,{\epsilon^\mu}_{\nu\alpha\beta}$,
the cosmological constant drops out completely due to the presence of $k^\nu k^\beta$.

One can first decompose the MPD equation of motion directly in the NP tetrad, while employing the Weyl-scalar relations summarized in paper I, equations (A1)--(A4):
\begin{align}
  k_\mu\dot{p}^\mu &= -s\,{^*\!}C_{\mu\nu\alpha\beta}k^\mu k^\nu l^\alpha k^\beta =0 \,,
                      \label{k.dotp} \\
  l_\mu\dot{p}^\mu &= -s\,{^*\!}C_{\mu\nu\alpha\beta}l^\mu k^\nu l^\alpha k^\beta
                    = 2s\,{\rm Im}\Psi_2 \,,  \label{l.dotp} \\
  m_\mu\dot{p}^\mu &= -s\,{^*\!}C_{\mu\nu\alpha\beta}m^\mu k^\nu l^\alpha k^\beta
                    = -{\rm i}s\Psi_1 \,,  \label{m.dotp} \\
  \bar{m}_\mu\dot{p}^\mu
                   &= -s\,{^*\!}C_{\mu\nu\alpha\beta}\bar{m}^\mu k^\nu l^\alpha k^\beta
                    =  {\rm i}s\overline{\Psi}_1 \,.  \label{barm.dotp}
\end{align}
It may however be more natural to escape the complex results by writing the last two components as projected onto the (real) orthonormal vectors (\ref{e2,e3}) rather than onto their complex null counter-parts. Since
\[e^\mu_{(2)}=\frac{1}{\sqrt{2}}\,(m^\mu+\bar{m}^\mu),  \quad
  e^\mu_{(3)}=\frac{1}{\sqrt{2}\,{\rm i}}\,(m^\mu-\bar{m}^\mu),\]
we find easily, in lieu of (\ref{m.dotp}) and (\ref{barm.dotp}),
\begin{align}
  e^{(2)}_\mu\dot{p}^\mu &= \sqrt{2}\,s\,{\rm Im}\Psi_1 \,,  \label{e2,dotp,m=0,ortho} \\
  e^{(3)}_\mu\dot{p}^\mu &= -\sqrt{2}\,s\,{\rm Re}\Psi_1 \,.  \label{e3,dotp,m=0,ortho}
\end{align}
In order to parallel the decomposition made in the massive case, one can also introduce orthonormal vectors
\[V^\mu=\frac{1}{\sqrt{2}}\,(k^\mu+l^\mu),  \quad
  e^\mu_{(1)}=\frac{1}{\sqrt{2}}\,(k^\mu-l^\mu)\]
and add the corresponding projections instead of (\ref{k.dotp}) and (\ref{l.dotp}),
\begin{equation}  \label{e01,dotp,m=0,ortho}
 -V_\mu\dot{p}^\mu = e^{(1)}_\mu\dot{p}^\mu = -\sqrt{2}\,s\,{\rm Im}\Psi_2 \,.
\end{equation}
The vector $V^\mu$ is a most natural time-like direction which the massless problem can be connected with; clearly, $e_{(1)}^\mu$ represents the corresponding spin vector (its unit form) -- it is orthogonal to $V^\mu$ and belongs to the spin-bivector eigen-plane ($S^{\mu\nu}e^{(1)}_\nu=0$).

Equations (\ref{e2,dotp,m=0,ortho}), (\ref{e3,dotp,m=0,ortho}) and (\ref{e01,dotp,m=0,ortho}) show that the projections of the massless pole-dipole MPD equation onto the ``natural" tetrad based on the world-line tangent $k^\mu$ are very simple and determined just by $\Psi_1$ and $\Psi_2$. In comparison with equations
\begin{eqnarray*}
  e^{\hat{1}}_\mu\,\dot{p}^\mu
     &=& -2s\,{\rm Im}\Psi_2,  \\
  e^{\hat{2}}_\mu\,\dot{p}^\mu
     &=& -s\,({\rm Im}\Psi_3-{\rm Im}\Psi_1), \\
  e^{\hat{3}}_\mu\,\dot{p}^\mu
     &=& -s\,({\rm Re}\Psi_3+{\rm Re}\Psi_1),
\end{eqnarray*}
obtained for massive particles and the Mathisson--Pirani spin condition (paper I), the massless case does not contain the $\Psi_3$ scalar. If one takes the advantage of the remaining freedom of the spin bivector subjected to only null spin condition $S^{\mu\nu}k_\nu=0$ (see section \ref{null-condition}) and {\em chooses} its second null eigen-direction $l^\mu$ to be given by the highest-multiplicity PND of the Weyl tensor (provided that $k_\mu l^\mu\neq 0$, of course), then, depending on the Petrov type, some of the Weyl scalars can be eliminated. In particular, besides $\Psi_4=0$ (note again that we take $l^\mu$ as the {\em second} vector of the NP tetrad), $\Psi_3$ / $\Psi_3$ and $\Psi_2$ / $\Psi_3$, $\Psi_2$ and $\Psi_1$ can thus be made vanish in type-II / type-III / type-N space-times. Hence, since the MPD-equation projections contain $\Psi_1$ and $\Psi_2$, they only simplify in type-III or type-N cases.

\subsection{Properties of the natural {\em orthonormal} tetrad}

Let us check some more properties of the above-introduced natural orthonormal tetrad
\begin{equation}  \label{ON-tetrad,massless}
  V^\mu, \quad
  e^\mu_{(1)}=:\frac{S^\mu}{s} \,, \quad
  e^\mu_{(2)}=\frac{p_\perp^\mu}{\cal M} \,, \quad
  e^\mu_{(3)}=\frac{s\,\dot{l}^\mu}{\cal M} \,.
\end{equation}
Firstly, provided that the particle's geodesic world-line is affinely parametrized, $\dot{k}^\mu=0$, we see that
\begin{equation}
  \dot{V}^\mu=\frac{\dot{l}^\mu}{\sqrt{2}}=\frac{{\cal M}\,e^\mu_{(3)}}{\sqrt{2}\,s} \;.
\end{equation}
One also easily relates the (null) spin $s^\mu$ to the newly introduced ``spin with respect to $V^\mu$" (denoted by $S^\mu$),
\begin{equation}
  s^\mu\equiv s\,k^\mu=\frac{s}{\sqrt{2}}(V^\mu+e_{(1)}^\mu)
                      =\frac{1}{\sqrt{2}}\left(sV^\mu+S^\mu\right).
\end{equation}
As $\dot{s}^\mu=s\,\dot{k}^\mu=0$, we have
\begin{equation}
  \dot{e}^\mu_{(1)}=-\dot{V}^\mu
                   =-\frac{{\cal M}\,e^\mu_{(3)}}{\sqrt{2}\,s} \;.
\end{equation}
Finally, regarding that
\begin{equation}
  \dot{p}_\perp^\mu=\dot{p}^\mu+k^\mu l_\nu\dot{p}^\nu
                   =(\delta^\mu_\nu+k^\mu l_\nu+l^\mu k_\nu)\,\dot{p}^\nu \,,
\end{equation}
one finds, from orthonormality of the basis,
\begin{align}
  \dot{e}^\mu_{(2)} &= \frac{e^{(3)}_\nu\dot{p}^\nu}{\cal M}\,e^\mu_{(3)} \,,  \label{dot(e2)} \\
  \dot{e}^\mu_{(3)} &= \frac{{\cal M}}{s}\,k^\mu
                       -\frac{e^{(3)}_\nu\dot{p}^\nu}{\cal M}\,e^\mu_{(2)} \,.
\end{align}

Having introduced $V^\mu$ and $S^\mu$, we can express the spin bivectors alternatively as\footnote
{Note that the above-introduced spin $S^\mu$ thus fixes the spin bivector {\em uniquely}, in contrast to the null spin $s^\mu$ introduced by (\ref{spin-vector}).}
\begin{equation}
  S_{\alpha\beta}=\epsilon_{\alpha\beta\gamma\delta}V^\gamma S^\delta, \quad
  {^*\!}S^{\mu\nu}=S^\mu V^\nu-V^\mu S^\nu
\end{equation}
and write, similarly as in paper I (section II.C), equations for $\dot{V}^\mu$ and $\dot{S}^\mu$ in terms of $k^\mu$ and $p^\mu$. Actually, multiplying
$\dot{S}_{\alpha\beta}=
 \epsilon_{\alpha\beta\gamma\delta}\dot{V}^\gamma S^\delta+
 \epsilon_{\alpha\beta\gamma\delta}V^\gamma\dot{S}^\delta$
by $\epsilon^{\mu\nu\alpha\beta}V_\nu$ and $\epsilon^{\mu\nu\alpha\beta}S_\nu$, we obtain, respectively,
\begin{align}
  \dot{S}^\mu &= \epsilon^{\mu\nu\alpha\beta} V_\nu k_\alpha p_\beta
               = -{^*\!}\dot{S}^{\mu\nu}V_\nu \,, \\
  s^2\dot{V}^\mu &= \epsilon^{\mu\nu\alpha\beta}S_\nu k_\alpha p_\beta
                  = -{^*\!}\dot{S}^{\mu\nu}S_\nu \,,
\end{align}
where we have already regarded that $s\dot{V}^\mu=-\dot{S}^\mu\sim e_{(3)}^\mu$ and $\dot{s}=0$.
Note that the above equations can also be obtained very straightforwardly by differentiating
\begin{equation}
  S^\mu = -{^*\!}S^{\mu\nu}V_\nu \,, \qquad
  s^2 V^\mu = -{^*\!}S^{\mu\nu}S_\nu \,,
\end{equation}
and that, thanks to $S_\mu S^\mu=s^2$, the magnitude of $S^\mu$ is automatically constant along $k^\mu$.

\subsection{Special cases of motion}

The massless spinning-particle problem turns out to be quite constrained, it offers much less freedom for various special performances than the massive case. Let us still mention two cases which arise naturally.

\subsubsection{The $p^\mu\sim k^\mu$ circumstance}
\label{p||k}

Notice, finally, that the tetrad (\ref{ON-tetrad,massless}) would be meaningless if $p^\mu$ belonged to the eigen-plane of $S^{\mu\nu}$ (i.e., if it were some space-like combination of $k^\mu$ and $l^\mu$), because then $s^2\dot{l}^\mu\equiv -S^{\mu\alpha}p_\alpha=0$ and, consequently, $\dot{S}_{\alpha\beta}=0$ and $p_\perp^\mu\equiv S^{\mu\nu}\dot{l}_\nu=0$.
In such a case, all the vectors $k^\mu$, $l^\mu$, $s^\mu$, $V^\mu$, $S^\mu$ and $p^\mu$ would lie in the spin-bivector eigen-plane, and most of them would be parallel transported along the representative world-line: $\dot{k}^\mu=0$, $\dot{l}^\mu=0$, $\dot{s}^\mu=0$, $\dot{V}^\mu=0$, $\dot{S}^\mu=0$. However, as already noted below equation (\ref{dotl.dotl}) and as best seen from equation (\ref{p-k,relation}), such a circumstance would imply $p^\alpha=-p^\beta l_\beta\,k^\alpha$, so ${\cal M}=0$ and $\dot{p}^\alpha=-\dot{p}^\beta l_\beta\,k^\alpha$, i.e. both $p^\alpha$ and $\dot{p}^\alpha$ would also have to be light-like and proportional to $k^\mu$. According to equation (\ref{*Cklk}), this would require
${^*\!C^\mu}_{\nu\alpha\beta}k^\nu l^\alpha k^\beta$ to be light-like, which definitely does not hold for {\em generic} motion in {\em generic} space-time. Using the metric decomposition
\begin{equation}  \label{metric-decomposition}
  g^{\mu\alpha}=-k^\mu l^\alpha-l^\mu k^\alpha+m^\mu\bar{m}^\alpha+\bar{m}^\mu m^\alpha
\end{equation}
and regarding that the first two terms yield zero in the scalar product below, one can rewrite the requirement as
\begin{align}
  0 &= ({^*\!}C_{\mu\nu\kappa\lambda}k^\nu l^\kappa k^\lambda)
       \,g^{\mu\alpha}\,
       ({^*\!}C_{\alpha\beta\gamma\delta}k^\beta l^\gamma k^\delta)=  \nonumber \\
    &=2\,({^*\!}C_{\mu\nu\kappa\lambda}m^\mu k^\nu l^\kappa k^\lambda)
         ({^*\!}C_{\alpha\beta\gamma\delta}\bar{m}^\alpha k^\beta l^\gamma k^\delta)=  \nonumber \\
    &=2\,\Psi_1\overline{\Psi}_1 \,,
\end{align}
which is only satisfied for $\Psi_1=0$, i.e., if i) either the particle moves in the direction ($k^\mu$) of the double PND of a Petrov-type-II space-time, ii) or the space-time is of type N (and one aligns with its quadruple PND the second vector $l^\mu$ of the NP tetrad).

\subsubsection{Stationary situation}

The only basic scalar involved which may not be constant is ${\cal M}$. Consider now the case when it {\em is} constant, $\dot{\cal M}=0$, but when ${\cal M}\neq 0$, so $p^\mu$ is space-like (if $p^\mu$ were light-like, it would immediately lead to $p^\mu\sim k^\mu$, which has already been mentioned above). From (\ref{MdotM}) one infers -- in both cases -- that $\dot{S}^{\alpha\beta}\dot{p}_\beta=0$, which implies that $\dot{p}^\mu$ belongs to the eigen-plane of $\dot{S}^{\mu\nu}$. This eigen-plane is spanned by $k^\mu$ and $\dot{l}^\mu$, so $\dot{p}^\mu$ has to be given by their combination, say $\dot{p}^\mu=\alpha k^\mu+{\cal M}\beta\,\dot{l}^\mu$. In particular, $\dot{p}^\mu_\perp$ must be proportional to $\dot{l}^\mu$, since it does not have any component proportional to $k^\mu$ by definition. Actually, the latter also follows, given $\dot{\cal M}=0$, from (\ref{dot(e2)}).\footnote
{Therefore, if $\dot{\cal M}=0$, then $\dot{p}^\mu_\perp$ can be used, after normalization, as the $e_{(3)}^\mu$ vector of the interpretation tetrad equally well as $\dot{l}^\mu$.}

A related consequence of $\dot{\cal M}=0$ is of course $p_\mu\dot{p}^\mu=0$. Writing $\dot{p}^\mu$ as (\ref{*Cklk}), inserting the metric (\ref{metric-decomposition}) and using $k_\sigma p^\sigma=0$ and
\[\dot{l}_\sigma p^\sigma=0 \quad \Rightarrow \quad
  m_\sigma p^\sigma=\bar{m}_\sigma p^\sigma=
  \frac{e^{(2)}_\sigma p^\sigma}{\sqrt{2}}=\frac{\cal M}{\sqrt{2}} \;,\]
one can express the $p_\mu\dot{p}^\mu=0$ circumstance as a simple condition for the type of space-time, because in terms of the Weyl scalars computed in our NP tetrad it reads
\begin{align}
  0 &= p_\mu\dot{p}^\mu
     = -s\,{^*\!C}_{\mu\nu\alpha\beta}\,p^\mu k^\nu l^\alpha k^\beta=  \nonumber \\
    &= -s\,{^*\!C}_{\mu\nu\alpha\beta}
       (m^\mu\bar{m}_\sigma+\bar{m}^\mu m_\sigma)\,p^\sigma k^\nu l^\alpha k^\beta=  \nonumber \\
    &= -\frac{{\cal M}s}{\sqrt{2}}\,{^*\!C}_{\mu\nu\alpha\beta}
       (m^\mu+\bar{m}^\mu)\,k^\nu l^\alpha k^\beta= \\
    &=\frac{{\cal M}s}{\sqrt{2}}\,(-{\rm i}\Psi_1+{\rm i}\overline{\Psi}_1)
     =\sqrt{2}{\cal M}s\,{\rm Im}\Psi_1 \,,
\end{align}
where, in the last row, equations (\ref{m.dotp}) and (\ref{barm.dotp}) have been used.

The coefficients of the $\dot{p}^\mu=\alpha k^\mu+{\cal M}\beta\,\dot{l}^\mu$ relation can also be found in terms of the  Weyl scalars: multiplying it by $l_\mu$ and $\dot{l}_\mu$, we have, respectively,
\begin{eqnarray}
  l_\mu\dot{p}^\mu=-\alpha &\;\dots\; =& 2s\,{\rm Im}\Psi_2 \,, \\
  \dot{l}_\mu\dot{p}^\mu={\cal M}\beta\,\frac{{\cal M}^2}{s^2}
                          &\;\dots\; =& -\sqrt{2}\,{\cal M}\,{\rm Re}\Psi_1 \,.
\end{eqnarray}

\section{Concluding remarks}

We have continued the study of a spinning-particle motion in the pole-dipole approximation. After treating, in \cite{SemerakS-PRDI}, the MPD equation of motion in an orthonormal tetrad tied to the ``reference observer" (denoted $V^\mu$), i.e. in a tetrad involving as time leg the vector which fixes the spin supplementary condition ($S^{\mu\nu}V_\nu=0$), we have considered the tetrad tied to the tangent of the world-line that represents the particle's history (denoted $u^\mu$). Both possibilities lead to usable formulations of the problem, with the latter (proposed in the present paper) being slightly less efficient, because $u^\mu$ cannot be freely chosen (in contrast to $V^\mu$). In both cases, we showed how the MPD equation decomposes if representing the curvature terms in the language of Weyl-tensor scalars obtained in the NP null tetrads naturally associated with the orthonormal ones. In the case of decomposing the MPD equation in the $u^\mu$-based tetrad, we have also shown how the projections look when computing the Weyl scalars in a {\em different} NP tetrad (different than the naturally associated with the orthonormal $u^\mu$-based tetrad), namely the one tied to a freely choiceable vector $V^\mu$.

Expressing the MPD-equation components in terms of the Weyl scalars, one can infer whether and how the exercise simplifies in particular Petrov types, provided that the NP tetrad can be aligned with the highest-multiplicity PND. Such an alignment is of course more problematic for the $u^\mu$-based tetrad (if one does not want to necessarily resort to the $S^{\mu\nu}u_\nu=0$ spin condition) which is much less flexible. Another item has been to see how the problem depends on the spin supplementary condition. We saw, in particular, that for the most advantageous option $u^\mu\parallel p^\mu$, the interpretation tetrads we had suggested (as given ``intrinsically" by geometry of the problem itself) were not available (two of their vectors turn zero), and suggested simple alternatives (which on the contrary do not work for the $S^{\mu\nu}u_\nu=0$ condition).

The second part of the present paper has been devoted to spinning particles with zero rest mass.
For them, the world-lines are null geodesics, the spin vector is also light-like (and proportional to the world-line tangent), the momentum is {\em space-like} (or null in a certain limit, which however only corresponds to a specific motion in type-II fields), and the Mathisson-Pirani spin condition follows necessarily. In spite of these important differences, a similar analysis can be performed as in the massive case, in particular, a certain ``intrinsic" interpretation tetrad can again be proposed. The respective decomposition of the MPD equation of motion is considerably simpler than in the massive case, it contains only $\Psi_1$ and $\Psi_2$ scalars and not the cosmological constant. Even (some of) these are eliminated in type-III or type-N space-times if the second null eigen-direction $l^\mu$ of the spin bivector is identified with the main PND of the background curvature (this is possible thanks to the less restricting nature of the null Mathisson-Pirani condition), not mentioning the case when the particle moves, at least at a given point, along a PND.

\begin{acknowledgments}
I am grateful for support from the grant GACR-14-10625S of the Czech Science Foundation.
I also thank Milan \v{S}r\'amek for a careful reading of the paper and for very helpful discussions.
\end{acknowledgments}

\bibliography{spinning-II.bib}

\providecommand{\noopsort}[1]{}\providecommand{\singleletter}[1]{#1}%
\begin{thebibliography}{7}%
\makeatletter
\providecommand \@ifxundefined [1]{%
 \@ifx{#1\undefined}
}%
\providecommand \@ifnum [1]{%
 \ifnum #1\expandafter \@firstoftwo
 \else \expandafter \@secondoftwo
 \fi
}%
\providecommand \@ifx [1]{%
 \ifx #1\expandafter \@firstoftwo
 \else \expandafter \@secondoftwo
 \fi
}%
\providecommand \natexlab [1]{#1}%
\providecommand \enquote  [1]{``#1''}%
\providecommand \bibnamefont  [1]{#1}%
\providecommand \bibfnamefont [1]{#1}%
\providecommand \citenamefont [1]{#1}%
\providecommand \href@noop [0]{\@secondoftwo}%
\providecommand \href [0]{\begingroup \@sanitize@url \@href}%
\providecommand \@href[1]{\@@startlink{#1}\@@href}%
\providecommand \@@href[1]{\endgroup#1\@@endlink}%
\providecommand \@sanitize@url [0]{\catcode `\\12\catcode `\$12\catcode
  `\&12\catcode `\#12\catcode `\^12\catcode `\_12\catcode `\%12\relax}%
\providecommand \@@startlink[1]{}%
\providecommand \@@endlink[0]{}%
\providecommand \url  [0]{\begingroup\@sanitize@url \@url }%
\providecommand \@url [1]{\endgroup\@href {#1}{\urlprefix }}%
\providecommand \urlprefix  [0]{URL }%
\providecommand \Eprint [0]{\href }%
\providecommand \doibase [0]{http://dx.doi.org/}%
\providecommand \selectlanguage [0]{\@gobble}%
\providecommand \bibinfo  [0]{\@secondoftwo}%
\providecommand \bibfield  [0]{\@secondoftwo}%
\providecommand \translation [1]{[#1]}%
\providecommand \BibitemOpen [0]{}%
\providecommand \bibitemStop [0]{}%
\providecommand \bibitemNoStop [0]{.\EOS\space}%
\providecommand \EOS [0]{\spacefactor3000\relax}%
\providecommand \BibitemShut  [1]{\csname bibitem#1\endcsname}%
\let\auto@bib@innerbib\@empty
\bibitem [{\citenamefont {Semer\'ak}\ and\ \citenamefont
  {\v{S}r\'amek}(2015)}]{SemerakS-PRDI}%
  \BibitemOpen
  \bibfield  {author} {\bibinfo {author} {\bibfnamefont {O.}~\bibnamefont
  {Semer\'ak}}\ and\ \bibinfo {author} {\bibfnamefont {M.}~\bibnamefont
  {\v{S}r\'amek}},\ }\bibfield  {title} {\enquote {\bibinfo {title} {Spinning
  particles in vacuum spacetimes of different curvature types},}\ }\href@noop
  {} {\bibfield  {journal} {\bibinfo  {journal} {Phys. Rev. D}\ }\textbf
  {\bibinfo {volume} {92}},\ \bibinfo {pages} {064032} (\bibinfo {year}
  {2015})}\BibitemShut {NoStop}%
\bibitem [{\citenamefont {Hall}(2004)}]{Hall-04}%
  \BibitemOpen
  \bibfield  {author} {\bibinfo {author} {\bibfnamefont {G.~S.}\ \bibnamefont
  {Hall}},\ }\href@noop {} {\emph {\bibinfo {title} {Symmetries and Curvature
  Structure in General Relativity}}}\ (\bibinfo  {publisher} {World
  Scientific},\ \bibinfo {address} {Singapore},\ \bibinfo {year}
  {2004})\BibitemShut {NoStop}%
\bibitem [{\citenamefont {Bailyn}\ and\ \citenamefont
  {Ragusa}(1977)}]{BailynR-77}%
  \BibitemOpen
  \bibfield  {author} {\bibinfo {author} {\bibfnamefont {M.}~\bibnamefont
  {Bailyn}}\ and\ \bibinfo {author} {\bibfnamefont {S.}~\bibnamefont
  {Ragusa}},\ }\bibfield  {title} {\enquote {\bibinfo {title} {Pole-dipole
  model of massless particles},}\ }\href@noop {} {\bibfield  {journal}
  {\bibinfo  {journal} {Phys. Rev. D}\ }\textbf {\bibinfo {volume} {15}},\
  \bibinfo {pages} {3543} (\bibinfo {year} {1977})}\BibitemShut {NoStop}%
\bibitem [{\citenamefont {Bailyn}\ and\ \citenamefont
  {Ragusa}(1981)}]{BailynR-81}%
  \BibitemOpen
  \bibfield  {author} {\bibinfo {author} {\bibfnamefont {M.}~\bibnamefont
  {Bailyn}}\ and\ \bibinfo {author} {\bibfnamefont {S.}~\bibnamefont
  {Ragusa}},\ }\bibfield  {title} {\enquote {\bibinfo {title} {Pole-dipole
  model of massless particles. {II}},}\ }\href@noop {} {\bibfield  {journal}
  {\bibinfo  {journal} {Phys. Rev. D}\ }\textbf {\bibinfo {volume} {23}},\
  \bibinfo {pages} {1258} (\bibinfo {year} {1981})}\BibitemShut {NoStop}%
\bibitem [{\citenamefont {Duval}\ and\ \citenamefont
  {Fliche}(1978)}]{DuvalF-78}%
  \BibitemOpen
  \bibfield  {author} {\bibinfo {author} {\bibfnamefont {C.}~\bibnamefont
  {Duval}}\ and\ \bibinfo {author} {\bibfnamefont {H.~H.}\ \bibnamefont
  {Fliche}},\ }\bibfield  {title} {\enquote {\bibinfo {title} {A conformal
  invariant model of localized spinning test particles},}\ }\href@noop {}
  {\bibfield  {journal} {\bibinfo  {journal} {J. Math. Phys.}\ }\textbf
  {\bibinfo {volume} {19}},\ \bibinfo {pages} {749} (\bibinfo {year}
  {1978})}\BibitemShut {NoStop}%
\bibitem [{\citenamefont {Mashhoon}(1975)}]{Mashhoon-75}%
  \BibitemOpen
  \bibfield  {author} {\bibinfo {author} {\bibfnamefont {B.}~\bibnamefont
  {Mashhoon}},\ }\bibfield  {title} {\enquote {\bibinfo {title} {Massless
  spinning test particles in a gravitational field},}\ }\href@noop {}
  {\bibfield  {journal} {\bibinfo  {journal} {Annals Phys.}\ }\textbf {\bibinfo
  {volume} {89}},\ \bibinfo {pages} {254} (\bibinfo {year} {1975})}\BibitemShut
  {NoStop}%
\bibitem [{\citenamefont {Bini}\ \emph {et~al.}(2006)\citenamefont {Bini},
  \citenamefont {Cherubini}, \citenamefont {Geralico},\ and\ \citenamefont
  {Jantzen}}]{BiniCGJ-06}%
  \BibitemOpen
  \bibfield  {author} {\bibinfo {author} {\bibfnamefont {D.}~\bibnamefont
  {Bini}}, \bibinfo {author} {\bibfnamefont {C.}~\bibnamefont {Cherubini}},
  \bibinfo {author} {\bibfnamefont {A.}~\bibnamefont {Geralico}}, \ and\
  \bibinfo {author} {\bibfnamefont {R.~T.}\ \bibnamefont {Jantzen}},\
  }\bibfield  {title} {\enquote {\bibinfo {title} {Massless spinning test
  particles in algebraically special vacuum space-times},}\ }\href@noop {}
  {\bibfield  {journal} {\bibinfo  {journal} {Int. J. Mod. Phys. D}\ }\textbf
  {\bibinfo {volume} {15}},\ \bibinfo {pages} {737} (\bibinfo {year}
  {2006})}\BibitemShut {NoStop}%
\end{thebibliography}%

\end{document}